\documentclass[aps,preprint]{revtex4}
\usepackage{amssymb,epsf}
\usepackage{amsmath}
\usepackage{graphicx}
\usepackage{hyperref}

\begin{document}

\title{Lifshitz Quartic Quasitopological Black Holes}
\author{M. Ghanaatian$^{1}$, A. Bazrafshan$^{2}$, and W. G. Brenna$^3$\thanks{%
email address: wbrenna@uwaterloo.ca}}
\affiliation{$^1$Department
of Physics, Payame Noor University, Iran}
\affiliation{$^2$Department of Physics, Jahrom University,
74137-66171 Jahrom, Iran}
\affiliation{$^3$Department of Physics
and Astronomy, University of Waterloo, Waterloo, Ontario N2L 3G1,
Canada}

\begin{abstract}
    In this paper we shall elucidate some of the effects of the
    quartic quasitopological term for Lifshitz-symmetric black holes. The field
    equations of this theory are difficult to
    solve exactly; here we will use numerical solutions both to verify
    previous exact solutions for quartic quasitopological AdS black holes as
    well as to examine new quasitopological Lifshitz-symmetric black hole solutions, in
    order to determine the effect of the quartic coupling parameter on the
    black hole's thermodynamic behaviour.
    We shall find that the quartic parameter controls solutions very similarly to
    the cubic parameter, allowing for the construction of a theory with another
    free parameter which may find meaning in the phase transition behaviour of a gauge/gravity context.
\end{abstract}

 \maketitle

\section{Introduction}
It is suspected that quantum gravity can be explained by a topological field theory, in
the sense that all of the gravitational degrees of freedom live on the boundary field theory.
This is the principle of holography; much progress in the
understanding of the holographic principle has been made in recent
years. The evidence for holography has been
explored since 1997 when Juan Maldacena conjectured the AdS/CFT
correspondence \cite{Maldacena1,Maldacena2}.

The AdS/CFT correspondence relates an asymptotically anti-de
Sitter (AdS) bulk theory with gravity in $(n+1)$-dimensions to a
conformal field theory on its $n$-dimensional Minkowski spacetime
boundary at infinity. As Einstein gravity does not have enough free
parameters to make a one-to-one relationship between
central charges and couplings on the non-gravitational side and the
``coupling'' parameters on the gravitational side, one may wish
to study modified theories, such as Lovelock theory
\cite{Friedman1,Friedman2,Jacobson} or quasitopological gravity
\cite{Lemos1,Lemos2,Lemos3,Aminneborg,Mann,Cai,Dehghani1,Dehghani2,Dehghani3,
Brenna2012},
where higher curvature terms produce additional coupling
parameters.

One could consider modifying the gravitational part of the Einstein
action with higher-derivative terms which arise in additional powers of
the curvature. Some gravity theories, like Lovelock gravity, play a
important role on the gravity side of the duality conjecture
\cite{Boera,Camanho}. The quasitopological framework is very similar to
Lovelock gravity, allowing for additional coupling
parameters in a given dimension, but at the cost of requiring spherical symmetry (otherwise the
equations of motion become greater than second-order).
The benefit to this approach is that quasitopological gravity can produce coupling
terms in fewer dimensions than Lovelock gravity, due to the fact that while Lovelock
terms become topological surface terms for a given number of dimensions, quasitopological terms are not true topological
invariants and therefore produce nontrivial gravitational effects in fewer dimensions than the corresponding Lovelock
terms.
Black hole solutions in quartic quasitopological gravity are not new \cite{Ghanaatian1,Ghanaatian2,Ghanaatian3}, but
the field equations generated by the quartic terms are lengthy and
numerical solutions are challenging to obtain.
Here we investigate asymptotically Lifshitz solutions of black holes
in quartic quasitopological gravity. The following (asymptotic)
form of spacetime metric is suggested from the holographically Lifshitz
condensed matter theories as
\begin{equation}
ds^{2}=L^{2}(-r^{2z}dt^{2}+\frac{dr^{2}}{r^{2}}+r^{2}dX^{2})
\end{equation}
which is based on the anisotropic scaling transformation (Lifshitz
scaling)
\begin{equation}
t\rightarrow{\lambda}^{z}t ,\text{ \ }
r\rightarrow{\lambda}^{-1}r,\text{ \ } X\rightarrow\lambda X
\end{equation}

This paper is organized as follows.
We review the action
of quartic quasitopological gravity and obtain the field
equations in Section \ref{sect:fe}. Section \ref{sect:cq} contains the calculation
of the conserved quantity along the radial coordinate $r$. In Section
\ref{sect:ls}, we consider two cases: the first Lifshitz solution is obtained
in the absence of matter, and then we derive the conditions on a Lifshitz solution
in the presence of a massive gauge field. Section \ref{sect:lsexp} is devoted to
the calculation of the asymptotically Lifshitz black hole
solutions near the horizon and at large $r$.
In Section \ref{sect:lscheck}, we first
check that new numerical solutions agree with exact solutions for
$z=1$, and then we find numerical quasitopological Lifshitz-symmetric black hole solutions
for $z=2$. In Section \ref{sect:lstd}, by
using the numerical results, we obtain the entropy and temperature
of the black hole and examine the thermodynamical behavior of the
black hole for $z=2$ in $3^{rd}$ and $4^{th}$ order
quasitopological gravity. Finally, we finish this paper with some
concluding remarks.

\section{The Field Equations Of Quasitopological Gravity}
\label{sect:fe}

Here we will derive field equations from the action of
quasitopological gravity up to the $4^{th}$ order. In addition, to
maintaining an asymptotically Lifshitz metric, a Proca field must be
introduced. The action can be written as
\begin{equation}
I = \int d^{n+1} x \sqrt{-g} \left(-2\Lambda +\mathcal{L}%
_{1}+{\mu _{2}}\mathcal{L}_{2}+{\mu _{3}}\mathcal{X}_{3}+{\mu _{4}}\mathcal{X}_{4}%
- \frac{1}{4}F_{\mu \nu}F^{\mu \nu} -\frac{1}{2}m^2A_{\mu}A^{\mu}
\right) \label{action}
\end{equation}
 where $(n+1)$ is the number of dimensions of
the spacetime, $F_{\mu \nu }=\partial _{\mu }A_{\nu }-\partial
_{\nu} A_{\mu }$ is the Proca field strength, $A_{\mu }$
is the Proca vector potential, $\mathcal{L}_{1}={R}$ is the
Einstein-Hilbert Lagrangian, and
$\mathcal{L}_{2}=R_{abcd}{R}^{abcd}-4{R}_{ab}{R}^{ab}+{R}^{2}$ is
the Gauss-Bonnet Lagrangian, $\mathcal{X}_{3}$  and
$\mathcal{X}_{3}$ are third and fourth order of quasitopological
gravity, respectively, which can be written as
(\cite{Myers2010,Ghanaatian1}):

\begin{eqnarray}
\mathcal{X}_{3} &=&R_{ab}^{cd}R_{cd}^{\,\,e\,\,\,f}R_{e\,\,f}^{\,\,a\,\,\,b}+%
\frac{1}{(2n-1)(n-3)}\left(
\frac{3(3n-5)}{8}R_{abcd}R^{abcd}R\right.  \notag
\\
&&-3(n-1)R_{abcd}R^{abc}{}_{e}R^{de}+3(n+1)R_{abcd}R^{ac}R^{bd}  \notag \\
&&\left. +\,6(n-1)R_{a}{}^{b}R_{b}{}^{c}R_{c}{}^{a}-\frac{3(3n-1)}{2}%
R_{a}^{\,\,b}R_{b}^{\,\,a}R+\frac{3(n+1)}{8}R^{3}\right)
\end{eqnarray}

\begin{eqnarray}
\mathcal{X}_{4} &=&c_{1}R_{abcd}R^{cdef}R_{%
\phantom{hg}{ef}%
}^{hg}R_{hg}{}^{ab}+c_{2}R_{abcd}R^{abcd}R_{ef}R^{ef}+c_{3}RR_{ab}R^{ac}R_{c}{}^{b}+c_{4}(R_{abcd}R^{abcd})^{2}
\notag \\
&&\hspace{-0.1cm}%
+c_{5}R_{ab}R^{ac}R_{cd}R^{db}+c_{6}RR_{abcd}R^{ac}R^{db}+c_{7}R_{abcd}R^{ac}R^{be}R_{%
\phantom{d}{e}}^{d}+c_{8}R_{abcd}R^{acef}R_{\phantom{b}{e}}^{b}R_{%
\phantom{d}{f}}^{d}  \notag \\
&&\hspace{-0.1cm}%
+c_{9}R_{abcd}R^{ac}R_{ef}R^{bedf}+c_{10}R^{4}+c_{11}R^{2}R_{abcd}R^{abcd}+c_{12}R^{2}R_{ab}R^{ab}
\notag \\
&&\hspace{-0.1cm}%
+c_{13}R_{abcd}R^{abef}R_{ef}{}_{g}^{c}R^{dg}+c_{14}R_{abcd}R^{aecf}R_{gehf}R^{gbhd},
\label{X4}
\end{eqnarray}
where the $c_i$ (quasitopological fixed constants) have
dimensional dependence and are tuned as per \cite{Myers2010} to
produce a simplification of the action (seen below in equation
\ref{evalaction}). See \cite{Ghanaatian1} (equation (7)) for the detailed values
used here.

Note that $\mathcal{X}_{3}$ and $\mathcal{X}_{4}$ are only
effective in dimensions greater than four and they become trivial
in six and eight dimensions respectively (Refs.
\cite{Myers2010,Ghanaatian1}). We use the asymptotically Lifshitz metric in the
spherically symmetric case as follows:
\begin{equation}\label{metlif}
ds^2 = -\frac{r^{2 z}}{L^{2 z}} f(r) dt^2 + \frac{L^2 dr^2}{r^2
g(r)} +   r^2  d\Omega^2
\end{equation}
where boundary conditions require that $f(r)$ and  $g(r)$ must go
to 1 as $r$ goes to infinity. The term $d\Omega^2$ is the metric
of an $n+1$ dimensional hypersurface with constant curvature
 $(n-1)(n-2)k$ and volume $V_{n-1}$
\begin{equation}
d\Omega^2 =d{\theta_1}^2 + k^{-1}\sin^2 {\left(\sqrt{k} \theta_1\right)} \left( d{\theta_2}^2 + \displaystyle\sum\limits_{i=3}^{D-2%
    } \displaystyle\prod\limits_{j=2}^{i-1} \sin^2{\theta_j } d{\theta_i}^2 \right)
\end{equation}
where the parameter $k$ specifies hyperbolic, flat, and spherical
geometries with the values $-1$, $0$, or  $1$, respectively. For
$k=0$ a coordinate transformation will reduce this portion of the
metric to the form $\sum_{k}^{D-2}d{\theta_k}^2$. To match the
spherical symmetry, we use a radial gauge field ansatz; for
simplicity we extract the $r^z$ dependence and write it as
\begin{equation}\label{gfield}
A_t = q \frac{r^z}{L^z} h(r) .
\end{equation}
where $h(r)$ also tends to unity at $r \rightarrow \infty$.

Because we seek solutions in the context of a gauge/gravity duality,
we will examine a five dimensional gravity theory for applicability to
a four dimensional gauge theory.
Under these considerations, one can obtain
the effective action for the spherically symmetric case as:
\begin{eqnarray}
\label{evalaction}
I &=& \int d^{4} x \int dr \frac{ r^{z-1} }{k L^{z+1}} \sqrt{\frac{f}{g}} \left( \left\{ 3 r^4 \left( \frac{- \Lambda}{6}L^2 %
    - \Psi+ \hat{\mu_{2}} \Psi^2 + \hat{\mu_{3}} \Psi^3 +\hat{\mu_{4}} \Psi^4 \right)
    \right\}^{\prime} \right. \notag
\\
&& \left. +\frac{q^2 r^3}{2 f} \left( g \left( r h^{\prime} + z h
\right)^2 + m^2 L^2 h^2  \right) \right)
\end{eqnarray}
where $ \Psi = \left( g - \frac{L^2}{r^2} k\right) $
and the dimensionless parameters $%
\hat{\mu}_{2}$, $\hat{\mu}_{3}$ and $\hat{\mu}_{4}$ are
redefinitions of the dimensionless coupling constants (to produce
an action that is cleaner to vary):

\begin{equation*}
\hat{\mu}_{2}\equiv \frac{(n-2)(n-3)}{l^{2}}\mu _{2},\text{ \ \ \ }\hat{\mu}%
_{3}\equiv -\frac{(n-2)(n-5)(3n^{2}-9n+4)}{8(2n-1)l^{4}}\mu _{3},
\end{equation*}%
\begin{equation*}
\hat{\mu}_{4}\equiv {\frac{n\left( n-1\right) \left( n-2\right)
^{2}\left(
n-3\right) \left( n-7\right) ({{n}^{5}-15\,{n}^{4}+72\,{n}^{3}-156\,{n}%
^{2}+150\,n-42)}}{{l}^{6}}}\mu _{4},
\end{equation*}

Varying the action of equation (\ref{evalaction}) with respect
$g(r)$, $f(r)$, and $h(r)$ respectively yields the following
equations of motion:
\begin{align}
&\Lambda L^2 r^8+6\hat{ \mu_{4}} \left(1-2z \right)g^4 r^8 +
\left( 3z+3 \right) r^8 g - 6z \hat{\mu_{2}}r^8 g^2 +12 \hat{
\mu_{4}} L^2 k \left(3z-2 \right) g^3 r^6+ 6z \hat{\mu_{2}} r^6
L^2 k g \nonumber \\
&- 3 r^6 L^2 k - \left( 9z-3 \right)\hat{ \mu_{3}} r^8 g^3 +
\left( 18z-9 \right) \hat{\mu_{3}} r^6 L^2 k g^2 +36 \hat{ \mu_{4}} L^4 k^2 \left(1-z\right) r^4- \left(9z-9 \right) \hat{\mu_{3}} L^4 k^2 r^4 g\nonumber \\
&+12 \hat{ \mu_{4}} L^6 k \left(z-2 \right) g r^2- 3 \hat{\mu_{3}}
L^6 k^3 r^2+6 \hat{ \mu_{4}}L^8 k^2
+ g (\ln{f})^{'} \left( \frac{3}{2} r^9 - 3 \right. \hat{\mu_{2}} r^9 g + 3 \hat{\mu_{2}} r^7 L^2 k \nonumber \\
&- \frac{9}{2} \hat{\mu_{3}} r^9 g^2 + \left. 9 \hat{\mu_{3}} r^7
g L^2 k - \frac{9}{2} \hat{\mu_{3}} r^5 L^4 k^2 -6 \hat{ \mu_{4}}
g^3 r^9+18 \hat{ \mu_{4}}L62 k g^2 r^7-18 \hat{ \mu_{4}} L^4 k^2 g
r^5+6 \hat{ \mu_{4}} r^3 L^6 k  \right)
\nonumber \\
& \quad \quad \quad \quad = \frac{q^2 r^8}{4 f}\left[ g \left(r h^{'} + z h \right)^2 - m^2 L^2 h^2 \right] \label{eom1}\\
&\left( 3 r^4 \left[ - \frac{\Lambda}{6} L^2%
- \Psi + \hat{\mu_{2}} \Psi^2 + \hat{\mu_{3}} \Psi^3+ \hat{\mu_{4}} \Psi^4 \right] \right)^{\prime}%
= \frac{q^2 r^3}{2 f}\left[ g \left(r h^{'} + z h \right)^2 + m^2
L^2 h^2 \right]
\label{eom2}\\
&2r^{2}h^{\prime \prime }- r\left[ (\ln f)^{\prime }-(\ln%
g)^{\prime }\right]( rh^{\prime }+zh) +2(z+4) rh^{\prime }+6zh
=2m^{2}L^{2}\frac{h}{g} \label{eom3}
\end{align}
where a prime ($^\prime$) represents the derivative with respect
to the radial coordinate $r$.

\section{The Conserved Quantity}
\label{sect:cq}

In order to calculate the conserved quantity along the radial
coordinate $r$, we calculate the first integral of the equations
of motion. Since there is no exact quasitopological-Lifshitz
solution, we can evaluate the conserved quantity at $r=\infty$
and at the horizon in order to obtain it explicitly.

It is simpler if we redefine the metric using a different ansatz,
as in \cite{Dehghani2010}.
\begin{align}
F(r) &= \frac{1}{2} \ln{f(r)} + z \ln{\frac{r}{L}}, \nonumber \\
G(r) &= -\frac{1}{2} \ln{g(r)} - \ln{\frac{r}{L}}, \nonumber \\
R(r) &= \ln{\frac{r}{L}}, \nonumber \\
H(r) &= \ln{h(r)} + z \ln{\frac{r}{L}},
\end{align}
the metric may be written as:
\begin{equation}
ds^2 = -e^{2 F(r)} dt^2 + e^{2 G(r)} dr^2 + e^{2 R(r)}
\frac{1}{L^2} d\Omega^2
\end{equation}

One can reduce the action to one dimension and obtain the equation
of motion. We insert this into the action (\ref{evalaction});
after integrating by parts we obtain a one dimensional Lagrangian
$\mathcal{L}_{1D} = \mathcal{L}_{1g} + \mathcal{L}_{1m}$ where
\begin{align}
\mathcal{L}_{1g} &= (n-1) \left( -2 \frac{\hat{\mu_{2}}}{n-1}e^{2G} + \left[ 2 F^{'} R^{'}%
+ (n-1) R^{\prime 2} \right] \right. \nonumber \\
&- \frac{\hat{\mu_{2}} L^2}{3} \left[ 4F^{\prime} R^{\prime 3} +%
(n-4)R^{\prime 4}\right] e^{-2G} \nonumber \\
& \quad - \left.\vphantom{\frac{\hat{\mu_{2}}}{n-1}}%
\frac{\hat{\mu_{3}}}{5} L^4 \left[ 6F^{\prime}R^{\prime 5} + (n-6)R^{\prime 6} \right]%
e^{-4 G} \right. \nonumber \\
&\left. -\frac{\hat{\mu_{{4}}}}{7}{L}^{6}{e^{-6\,G}} \left(
8\,{\it F^{'}}\,{{\it R^{'}}
}^{7}+ \left( n-8 \right) {{\it R^{'}}}^{8} \right) \right) e^{F - G + (n-1) R} \\
\mathcal{L}_{1m} &= \frac{1}{2}q^2 \left( m^2 + H^{\prime 2} e^{-2G} \right)%
e^{-F + G + (n-1)R + 2H}.
\end{align}
Calculating the equations of motion from the above action, we
have:
\begin{align}\label{sublag}
\mathcal{L}_{1g}-\mathcal{L}_{1m} = \left\{2\left( n-1 \right)
{e^{F-G+ \left( n-2 \right) R}} \right. & \left. \left( \,{\it
R^{'}}-\frac{2}{3}\,\hat{\mu_{{2}}}{l}^{2}{e^{-2\,G}}{{\it
R^{'}}}^{3} \right. \right. \nonumber \\
& \left. \left. -\frac{3}{5}\,\hat{\mu_{{3}}}{l}^{4}{e^{-4\,G}}{{\it
R^{'}}}^{5}-{\frac {4}{7}}\,\hat{\mu_{{4}}}{l}^{
6}{e^{-6\,G}}{{\it R^{'}}}^{7} \right)\right\}^{'}
\end{align}
\begin{align}\label{sumlag}
\mathcal{L}_{1g}+\mathcal{L}_{1m} = \left\{{e^{F-G+ \left( n-1
\right) R}} \right. & \left. \left( 2\,{\it F^{'}}+2\, \left( n-1
 \right) {\it R^{'}}-\frac {1}{3}\,\mu_{{2}}{l}^{2}{e^{-2\,G}} \left( 12\,{
\it F^{'}}\,{{\it R^{'}}}^{2}+4\, \left( n-4 \right) {{\it
R^{'}}}^ {3} \right) \right. \right. \nonumber \\
& \left. \left. -\frac{1}{5}\,\mu_{{3}}{l}^{4}{e^{-4\,G}}
\left( 30\,{\it F^{'}} \,{{\it R^{'}}}^{4}+6\, \left( n-6 \right)
{{\it R^{'}}}^{5}
 \right) \right. \right. \nonumber \\
& \left. \left. -\frac {1}{7}\,\mu_{{4}}{l}^{6}{e^{-6\,G}} \left( 56\,{\it F^{'}}\,{{
\it R^{'}}}^{6}+8\, \left( n-8 \right) {{\it R^{'}}}^{7} \right)
 \right)\right\}^{'}
\end{align}
\begin{align}\label{2lagm}
2\mathcal{L}_{1m} &= \left\{{q}^{2}{\it H^{'}}\,{e^{-F-G+ \left(
n-1 \right) R+2\,H}}\right\}^{'}
\end{align}
We are able to obtain the conserved quantity by subtracting the
sum of Eq. (\ref{sublag}) and Eq. (\ref{sumlag}) from Eq.
(\ref{2lagm}) (obtaining a total derivative) and integrating:
\begin{align}
\mathcal{C}_0 &= 2\left(F^{\prime}-R^{\prime}\right) \left(1 - 2\hat{\mu_{2}} L^2 R^{\prime%
2}e^{-2G} - 3 \hat{\mu_{3}} L^4 R^{\prime 4}e^{-4G}- 4 \hat{\mu_{4}} L^6 R^{\prime 6}e^{-6G} \right) e^{F - G + (n-1)R} \nonumber \\
& \quad - q^2 H^{\prime}e^{-F -G +(n-1)R + 2H} \nonumber \\
&= \left[ \left( 1 - 2 \hat{\mu_{2}} g - 3 \hat{\mu_{3}} g^2- 4 \hat{\mu_{4}} g^3 \right) \left( r f^{\prime} + 2 \left( z - 1 \right) f \right) - q^2 \left( z h + r h^{\prime} \right) h%
\right] \frac{r^{z+n-1}}{L^{z+1}} \left( \frac{f}{g} \right)^{1/2}
 \label{Constant}
\end{align}
Note that for $z=1$, where  $f(r) = g(r)$, the conserved quantity
reduces to
\begin{equation*}
\mathcal{C}_0 = \frac{r^{n+1}}{L^2} \left( f - \hat{\mu_{2}} f^2 -
\hat{\mu_{3}} f^3 - \hat{\mu_{4}} f^4 \right)^{\prime}
\end{equation*}
which is known to be constant in fourth order quasitopological
gravity and it is proportional to the mass of the black hole.

\section{Lifshitz Solutions}
\label{sect:ls}

\subsection{Matter-free Solutions}

In quasitopological gravity in the absence of matter, by setting
$h(r)=0$, in 5 dimensions we investigate the  solutions of the
form
\begin{equation}
ds^{2}=-\frac{r^{2z}}{L^{2z}}dt^{2}+\frac{L^2 dr^{2}}{r^{2}}+r^{2}\sum%
\limits_{i=1}^{3}d\theta _{i}^{2},  \label{met2}
\end{equation}
where $k=0$.  In order to obtain an asymptotically Lifshitz
solution in fourth order quasitopological gravity, the following
constraints arise, for an arbitrary value of z:
\begin{equation}\label{Condition1}
\Lambda = -\frac{2}{{L}^{2}}({\hat{\mu_{{4}}}+2-\hat{\mu_{{2}}}}), \hspace{5mm}%
\hat{\mu_{3}} = -\frac{1}{3}(4\hat{\mu_{{4}}}-1+2\hat{\mu_{{2}}})
,
\end{equation}

Inserting $\hat{\mu_{3}}=\hat{\mu_{4}} = 0$, these constraints
reduce to those of five dimensional Gauss-Bonnet gravity
\cite{Dehghani2010}
\begin{equation}
\Lambda = -\frac{3}{L^2} \text{ \ \ and \ \ } \hat{\mu_{2}} =
\frac{1}{2}
\end{equation}.

Note that with the above constraints, the exact Lifshitz solution
(that is, $f(r) = g(r) =1$) is a solution of the field equations
for any value of $z$. Inserting the conditions (\ref{Condition1})
into Eq. (\ref{eom2}), we have
\begin{equation}\label{kapeq}
2-\hat{\mu_{2}}+\hat{\mu_{4}}-3\Psi+3 \hat{\mu_{2}} \Psi^2 +
(1-2\hat{\mu_{2}}-4\hat{\mu_{4}})\Psi^3+3\hat{\mu_{4}}
\Psi^4=\frac{C}{r^4}
\end{equation}
where $C$ is a constant of integration. For $C=0$ ($\Psi=1$), one
can obtain the following result:
\begin{equation}
g(r) = 1 + \frac{k L^2}{r^2}
\end{equation}
Choosing $f(r) = g(r)$
for $k = -1$ yields an event horizon, therefore
the metric
\begin{equation}\label{metbhexact}
ds^2 = -\frac{r^{2 z}}{L^{2 z}} \left(1 - \frac{L^2}{r^2}\right)
dt^2 + \frac{L^2 dr^2}{r^2 (1 -\frac{L^2}{r^2})} + r^2
d\Omega_{-1}^2
\end{equation}
is an exact black hole solution, precisely as was found in \cite{Brenna2011}.

For $z=1$, one can extend the solution found in \cite{Brenna2011}.
Choosing $f(r) = g(r)$, $h(r)=0$, the field equation
(\ref{eom3}) disappears, and the equations (\ref{eom1}) and
(\ref{eom2}) can be analytically solved.
The real general solutions in $(n+1)$ dimensions of Eq.
(\ref{eom2}) are

\begin{equation}
f(r)=k+\frac{r^{2}}{l^{2}}\left( \frac{\hat{\mu}_{3}}{4\hat{\mu}_{4}}+\frac{1%
}{2}R\pm \frac{1}{2}E\right) .  \label{F4}
\end{equation}
where

\begin{eqnarray}
R &=&\left( \frac{{\hat{\mu}_{3}}^{2}}{4{\hat{\mu}_{4}}^{2}}-\frac{2\hat{\mu}%
_{2}}{3\hat{\mu}_{4}}+\left( {\frac{D}{2}+\sqrt{\Delta }}\right)
^{1/3}+\left( {\frac{D}{2}-\sqrt{\Delta }}\right) ^{1/3}\right)
^{1/2},
\label{RR} \\
E &=&\left( \frac{3{\hat{\mu}_{3}}^{2}}{4{\hat{\mu}_{4}}^{2}}-\frac{2\hat{\mu%
}_{2}}{\hat{\mu}_{4}}-R^{2}-\frac{1}{4R}\left[ \frac{4\hat{\mu}_{2}\hat{\mu}%
_{3}}{{\hat{\mu}_{4}}^{2}}-\frac{8}{\hat{\mu}_{4}}-\frac{{\hat{\mu}_{3}}^{3}%
}{{\hat{\mu}_{4}}^{3}}\right] \right) ^{1/2}  \label{EE}
\end{eqnarray}
and
\begin{equation*} \Delta
=\frac{C^{3}}{27}+\frac{D^{2}}{4}
\end{equation*}
\begin{equation}
C={\frac{3\hat{\mu}_{3}-{\hat{\mu}_{2}}^{2}}{3{\hat{\mu}_{4}}^{2}}}-\,{\frac{%
4\kappa}{\hat{\mu}_{4}}}
\end{equation}
\begin{equation}
D={\frac{2}{27}}\,{\frac{{\hat{\mu}_{2}}^{3}}{{\hat{\mu}_{4}}^{3}}}-\frac{1}{%
3}\,\left( {\frac{\hat{\mu}_{3}}{{\hat{\mu}_{4}}^{2}}} + 8\,{\frac{\kappa}{%
\hat{\mu}_{4}}}\right) \frac{\hat{\mu}_{2}}{\hat{\mu}_{4}}+{\frac{{\hat{\mu}%
_{3}}^{2}\kappa}{{\hat{\mu}_{4}}^{3}}}+\frac{1}{{\hat{\mu}_{4}}^{2}}
\end{equation}
and
\begin{equation}
\kappa =-\frac{2\Lambda}{n(n-1)}L^2-\frac{m}{r^{n}},  \label{kap}
\end{equation}
Note that $m$ is an integration constant.

\subsection{Matter solutions}

In this section we consider another case: Lifshitz solutions in
the presence of a massive gauge field $A^{\mu}$. Setting $h(r)\neq
0$ the Lifshitz solution (\ref{met2})  can be asymptotically obtained by using
the following constraints
\begin{align}\label{lifshitzlovelockexact}
q^2 &= \frac{2 \left( z - 1 \right) \left( 1 - 2\hat{\mu_{2}} -
3\hat{\mu_{3}}-4\hat{\mu_{4}} \right)}{z} \\
m^2 &= \frac{(n-1) z}{L^2}    \\
\Lambda &= -\frac{1}{2L^2} \left[ ( 1 - 2\hat{\mu_{2}} - 3\hat{\mu_{3}}-4\hat{\mu_{4}} ) \right. \nonumber \\
    & \hspace{15mm} \left. ((z-1)^2+n(z-2)+n^2 ) +  %
n(n-1)(\hat{\mu_{2}} +2\hat{\mu_{3}}+3\hat{\mu_{4}}) \right] \\
\hat{\mu_{2}} &< \frac{1}{2} ( 1 - 3\hat{\mu_{3}}-4\hat{\mu_{4}} )
\end{align}
where the last constraint arises because we require $q^{2}>0$.
This yields
\begin{align}
-\frac{1}{2L^2}\left[(1-3\hat{\mu_{3}}\right. & -4\hat{\mu_{4}})((z-1)^2+n(z-2)+n^2
) \nonumber \\
\left. +n(n-1)(2\hat{\mu_{3}}+3\hat{\mu_{4}})\right] & \leq \Lambda \leq
-\frac{1}{4L^2}[n(n-1)(1+\hat{\mu_{3}}+2\hat{\mu_{4}})]
\end{align}

In the Lifshitz metrics where $z>1$, we are unable to find exact
solutions so we use numerical methods.

\section{Asymptotically Lifshitz Black Holes}
\label{sect:lsexp}

In this section, we search for asymptotically Lifshitz black hole
solutions near the horizon and at large $r$. The general numeric
procedure will be to construct a series solution near the horizon
in order to form boundary conditions for the numerical solver.
Then, the field equations (our set of nonlinear ODEs in
$f(r),g(r),h(r)$) are solved numerically such that they also agree
with the Lifshitz asymptotics at large $r$. This is done by
iteratively solving the ODEs using the free parameters in the
series solution (for $z=1$ there are two free parameters, $h_1$
and $f_1$) until the solutions converge to unity, within some
tolerance (typically we set the tolerance to $10^{-8}$).

\subsection{Series Solutions Near The Horizon In 5 Dimensions}

Requiring that the solutions $f(r)$ and $g(r)$
tend to zero linearly near the horizon $r=r_0$, we write a
near-horizon series expansion as

\begin{align}
f(r)
&=f_{1}\left\{(r-r_{0})+f_{2}(r-r_{0})^{2}+f_{3}(r-r_{0})^{3}+...\right\}, \nonumber \\
g(r)
&=g_{1}(r-r_{0})+g_{2}(r-r_{0})^{2}+g_{3}(r-r_{0})^{3}+..., \label{near-hor}\\
h(r)
&=f_{1}^{1/2}\left\{h_{0}+h_{1}(r-r_{0})+h_{2}(r-r_{0})^{2}+h_{3}(r-r_{0})^{3}+...\right\},
\nonumber
\end{align}

Inserting the ansatz into the equations of the motion, we find
$h_0 = 0$ and the following  restriction on $g_1$:
\begin{align}
g_{1} &=\frac{z}{r_{0}^{3}}\left\{ \left(  \left( 4\,{z}^{2}+8\,z
\right) {{\it r_{0}}}^{8}+12\,{l}^{8}{k}^ {4} \right)\hat{
\mu_{{4}}}+ \left(  \left( 3\,{z}^{2}+6\,z+3 \right) {{\it
r_{0}}}^{8}-6\,{l}^{6}{k}^{3}{{\it r_{0}}}^{2} \right)
\hat{\mu_{{3}}} \right. \nonumber \\
& \left.  +\left( 2\, {z}^{2}+4\,z+6 \right) {{\it
r_{0}}}^{8}\hat{\mu_{{2}}}+ \left( -{z}^{2}-9-2\,z
 \right) {{\it r_{0}}}^{8}-6\,{l}^{2}k{{\it r_{0}}}^{6}
\right\} \nonumber \\
& \left\{ \left(  \left( -4\,z{{\it h_{1}}}^{2}+4\,{{\it
h_{1}}}^{2} \right) {{\it r_{0} }}^{7}+12\,z{l}^{6}{k}^{3} \right)
\hat{\mu_{{4}}}+ \left( \left( -3\,z{{ \it h_{1}}}^{2}+3\,{{\it
h_{1}}}^{2} \right) {{\it r_{0}}}^{7}-9\,{l}^{4}{k}^{2 }{{\it
r_{0}}}^{2}z \right) \hat{\mu_{{3}}} \right.  \nonumber \\
& \left.  + \left(  \left( -2\,z{{\it
h_{1}}}^{2} +2\,{{\it h_{1}}}^{2} \right) {{\it
r_{0}}}^{7}+6\,{l}^{2}k{{\it r_{0}}}^{4}z
 \right) \hat{\mu_{{2}}}+ \left( z{{\it h_{1}}}^{2}-{{\it h_{1}}}^{2} \right) {{
\it r_{0}}}^{7}+3\,{{\it r_{0}}}^{6}z \right\}^{-1}
\end{align}
where higher order expansion terms can also be written in terms of just $f_1,h_1$,
for example,
\begin{align}
g_{2} &=-{{\it r_{0}}}^{2} \left\{  (  (  ( 24\,\hat{\mu_{{3}}}+ %
32\,\hat{\mu_{{4}}}+16\,\hat{\mu_{{2}}}-8 ) {\it g_{1}}\,{{\it %
h_{1}}}^{2}-6\,\hat{\mu _{{2}}}{{\it g_{1}}}^{2})z \right. \nonumber
\\ & +(8-32\,\hat{\mu_{{4}}}-16\,\hat{\mu_{{2}}}- 24\,\hat{\mu_{{3}}} %
) {\it g_{1}}\,{{\it h_{1}}}^{2} ) {{\it r_{0}}}^{7}+ ( ( (%
-3+6\,\hat{\mu_{{2}}}+12\,\hat{\mu_{{4}}}+9\,\hat{\mu_{{3}}}
) {{\it h_{1}}}^{2}+12\,{\it g_{1}} ) {z}^{2} \nonumber
\\ &
+ ( -6\,\hat{\mu_
{{2}}}-9\,\hat{\mu_{{3}}}-12\,\hat{\mu_{{4}}}+3 ) {{\it
h_{1}}}^{2}z ) {{ \it r_{0}}}^{6}+18\,{{\it r_{0}}}^{5}{{\it
g_{1}}}^{2}z\hat{\mu_{{3}}}{l}^{2}k+
 ( 24\,{\it g_{1}}\,{z}^{2}\hat{\mu_{{2}}}{l}^{2}k-12\,z\hat{\mu_{{2}}}{l}^{2}k
{\it g_{1}} ) {{\it r_{0}}}^{4} \nonumber
\\ & \left.
-36\,{{\it r_{0}}}^{3}{{\it
g_{1}}}^{2}z\hat{\mu _{{4}}}{l}^{4}{k}^{2}+ ( 36\,{\it
g_{1}}\,z\hat{\mu_{{3}}}{l}^{4}{k}^{2}- 36\,{\it
g_{1}}\,{z}^{2}\hat{\mu_{{3}}}{l}^{4}{k}^{2} ) {{\it r_{0}}}^{2}+
48\,{\it g_{1}}\,{z}^{2}\hat{\mu_{{4}}}{l}^{6}{k}^{3}-72\,{\it
g_{1}}\,z\hat{\mu_{{4}} }{l}^{6}{k}^{3} \right\}
\nonumber
\\ &
\left.[ ( (
1-2\,\hat{\mu_{{2}}}-3\,\hat{\mu_{{3}}}- 4\,\hat{\mu_{{4}}} ) {\it
g_{1}}\,{{\it h_{1}}}^{2}z+ ( -1+2\,\hat{\mu_{{2}}
}+3\,\hat{\mu_{{3}}}+4\,\hat{\mu_{{4}}} ) {\it g_{1}}\,{{\it
h_{1}}}^{2} ) {{\it r_{0}}}^{10}+9\,{{\it r_{0}}}^{9}z{\it g_{1}}
\nonumber
\right.
\\ &
+( ( -1+2\,\hat{\mu_ {{2}}}+3\,\hat{\mu_{{3}}}+4\,\hat{\mu_{{4}}} )
{z}^{3}+ ( 8\,\hat{\mu_{{4}}}-2
+4\,\hat{\mu_{{2}}}+6\,\hat{\mu_{{3}}} ) {z}^{2}+ (
3\,\hat{\mu_{{3}}}-9+6\, \hat{\mu_{{2}}} ) z ) {{\it r_{0}}}^{8}
\nonumber
\\ & \left.
+18\,z\hat{\mu_{{2}}}{l}^{2}k{{ \it r_{0}}}^{7}{\it
g_{1}}-6\,z{l}^{2}k{{\it r_{0}}}^{6}-27\,z\hat{\mu_{{3}}}{l}^{4}
{k}^{2}{{\it r_{0}}}^{5}{\it
g_{1}}+36\,z\hat{\mu_{{4}}}{l}^{6}{k}^{3}{{\it r_{0}}}^ {3}{\it
g_{1}}-6\,z\hat{\mu_{{3}}}{{\it
r_{0}}}^{2}{l}^{6}{k}^{3}+12\,z\hat{\mu_{{4}}}{
l}^{8}{k}^{4}\right]^{-1}
\end{align}

From this method, we find that $h_1$ and $f_1$ are both free
parameters and should be chosen suitably to ensure proper
asymptotic behaviour for large $r$. We can derive the remaining
coefficients of the near horizon series solution up to third order
in terms of these free parameters.

What we find, consistent with the third-order
quasitopological case \cite{Brenna2011}, is that in $z=1$, both
parameters $h_1$ and $f_1$ must be tuned to guarantee the correct
asymptotic behaviour. However, in Lifshitz spacetimes, such as for
$z=2$, there are numerous values of $h_1$
which furnish a family of black hole solutions.
Furthermore, a similar approach to \cite{Brenna2011} yields the
existence of a valid large-$r$ series.
Therefore, solutions can have both a valid near-horizon and far-horizon
expansion.

\section{Numerical Solutions}
\label{sect:lscheck}

\subsection{Consistency Check}
\begin{figure}[htbp]
    \includegraphics[scale=0.5]{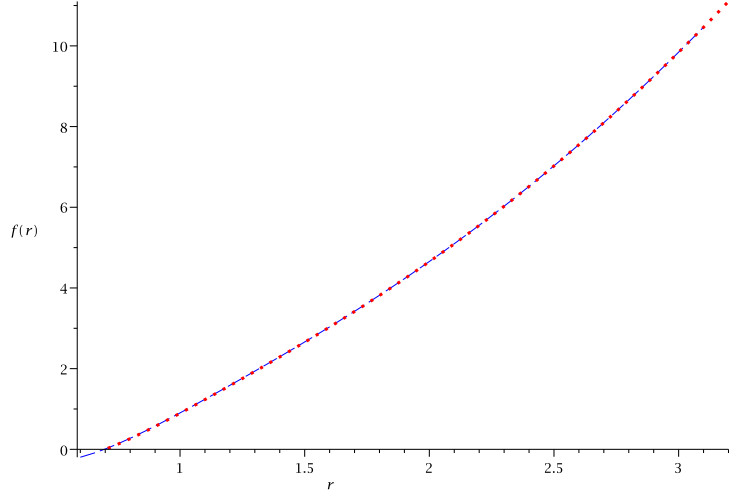}
    \caption{The overlay plot of $f(r)$ versus $r$ for the analytic solution
    (blue, dashed) and the numeric solution (red, dotted). Here, $\hat{\mu}_2 =%
    0.2$, $r_+ = 0.7$, $z=1$. The other parameters were
    $\hat{\mu}_0=0.817185,\hat{\mu}_3=-0.01778,\hat{\mu}_4=0.000593,m=0.9165$.}
    \label{fig:check}
\end{figure}

We begin the numerical procedure by ensuring that our solutions agree with previous
results. Due to the fact that the Lifshitz metric simplifies to
AdS when $z=1$, we attempted to replicate the exact solution from
\cite{Ghanaatian1}, specified in their equation (21). Because
their solution was not tuned to converge to unity at infinity, we
need to adjust the form of $\hat{\mu}_0$. Using equation (\ref{F4}),
and the fact that our definition of $\hat{\mu}_3$ is equal to the
negative value of their definition, this amounts to comparing our
numerical solution to the exact solution
\begin{equation}
    f(r) = k + \frac{3}{2\hat{\mu}_2}\frac{r^2}{l^2} \left( 1 - \left( 1 - %
    \frac{8\hat{\mu}_2}{3} \left[ \hat{\mu}_0 - \frac{m}{r^4} \right]%
    \right)^{1/4} \right)
    \label{eqn:exactdehghani}
\end{equation}
where we have the additional conditions
\begin{align*}
    \hat{\mu}_0 &= \frac{1}{2} \left(  ( 1 - 2\hat{\mu_{2}} + %
    3\hat{\mu_{3}}-4\hat{\mu_{4}} )\right. \nonumber \\
    & \left. \left( (z-1)^2+4(z-2)+16 ) +  %
    12 (\hat{\mu_{2}} -2\hat{\mu_{3}}+3\hat{\mu_{4}}) \right) \right) \\
    \hat{\mu}_3 &= -\frac{4 \hat{\mu}_2^2}{9} \\
    \hat{\mu}_4 &= \frac{2 \hat{\mu}_2^3}{27}
\end{align*}
and have specified $l=1$. Note that in their metric, the
$r^{2z} = r^2$ was absorbed into the definition of $f(r)$, so we
had to multiply our results by $r^2$. The geometric mass $m$
controls where the event horizon is located in the analytic
solution; we varied $m$ until it matched the value of the root we
chose: $r_+ = 0.7$. This corresponds to $m=0.9165$. We plot the
comparison for a value of $\hat{\mu}_2 = 0.2$ in figure
\ref{fig:check}.

\subsection{Behaviour of $f(r)$}
\begin{figure}[htbp]
    \includegraphics[scale=0.5]{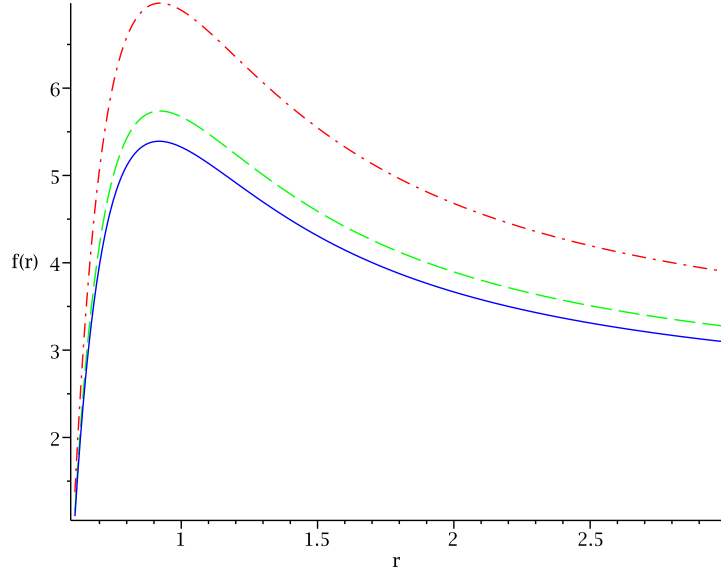}
    \caption{The plot of $f(r)$ versus $r$ for the different solutions
    in Gauss-Bonnet (blue, solid), $3^{rd}$-order quasitopological gravity
    (green, dashed), and $4^{th}$-order quasitopological gravity (red, dot-dash).
    Here, $r_+ = 0.59$, $k=1$, and $z=2$.
    The other parameters were
    $\hat{\mu}_2=0.2,\hat{\mu}_3=-0.001,\hat{\mu}_4=0.0005$.}
    \label{fig:z2qt4fr}
\end{figure}

Now that we have ensured accuracy of our results as compared to the exact
solutions, we are able to explore the effect of different values
of $z$ on the solutions. As an example, we examine the case where
$z=2$, using the parameters
$\hat{\mu}_2=0.2,\hat{\mu}_3=-0.001,\hat{\mu}_4=0.0005,r_+=0.59$,
where $k=1$ and the seed value for $h_0$ was 2.5. This is plotted in
figure \ref{fig:z2qt4fr}.

As we can see, a negative $3^{rd}$-order quasitopological
parameter makes the peak of $f(r)$ heightened, while a positive
$4^{th}$-order quasitopological parameter acts in the same manner
as the $3^{rd}$-order parameter.

\begin{figure}[htbp]
    \includegraphics[scale=0.5]{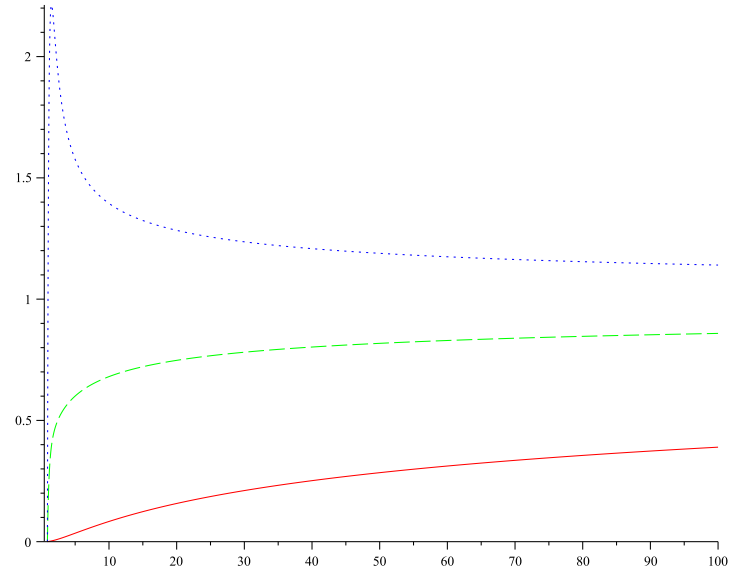}
    \caption{The plot of $f(r)$ versus $r$ for the different solutions
    in $4^{th}$-order quasitopological gravity for values of $k=1,0,-1$
    (dot, dash, and solid respectively).
    Here, $r_+ = 0.9$, and $z=2$.
    The other parameters were
    $\hat{\mu}_2=0.04,\hat{\mu}_3=-0.001,\hat{\mu}_4=0.0003$.}
    \label{fig:kvarfr}
\end{figure}

We can explore the effect of the hypersurface curvature on the solutions of
$f(r)$ by plotting a quasitopological numerical solution for values of
$k=1,0,-1$. We do this in figure \ref{fig:kvarfr}.

In addition, we can explore the family of black hole solutions
which arises as a result of the degeneracy in the Proca field
boundary conditions \cite{Andrade2012,Keeler2012}. As mentioned in
\cite{Brenna2011}, we have a set of black holes in which the
metric functions differ for different ``seed'' values of the
parameter $h_0$. We can explore how the $4^{th}$-order
quasitopological parameter will alter the behaviour of this family
of solutions.

\begin{figure}[htbp]
    \includegraphics[scale=0.5]{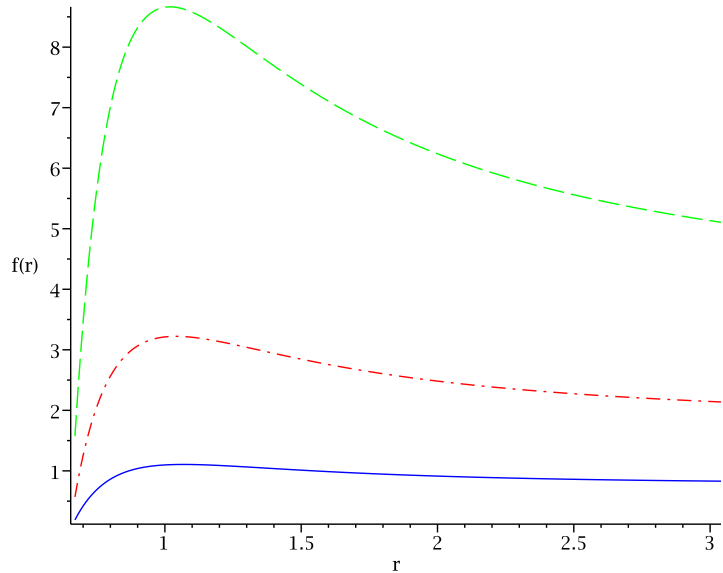}
    \caption{The plot of $f(r)$ versus $r$ for the different values
    of $h_0$: 2.5 (blue, solid), 2.3 (green, dashed), and 2.4 (red, dot-dash).
    Here, $r_+ = 0.65$, $k=1$, and $z=2$.
    The other parameters were
    $\hat{\mu}_2=0.2,\hat{\mu}_3=-0.001,\hat{\mu}_4=0.0005$  .}
    \label{fig:z2h0fr}
\end{figure}

Looking at a set of seed values in the range $h_0 = \left[ 2.3, 2.5
\right]$, we see in figure \ref{fig:z2h0fr} that the family of
solutions still exists in quartic quasitopological gravity.
The behaviour of the seed value also acts similarly; it allows some
control over the initial spike of $f(r)$.

Overall an examination of the metric functions shows that the
quartic quasitopological term, for the values studied, does not
cause dramatically different solutions when compared to those obtained from cubic
quasitopological gravity. This is not a disappointing conclusion;
the ability to add a new parameter to the black hole without much
additional cost could be important with respect to the gauge/gravity theory.

\section{Thermodynamics}
\label{sect:lstd}

The entropy of the black hole solutions can be calculated through
the use of the Iyer/Wald formula as \cite{Iyer}
\begin{equation}
S = -2 \pi \oint d^{n-1} x \sqrt{\tilde{g}} Y^{a b c d} \hat{\epsilon}_{ab} \hat{\epsilon}_{cd}, \hspace{5mm} \text{where} \hspace{2mm} %
Y^{a b c d} = \frac{\partial{\mathcal{L}}}{\partial{R_{a b c d}}}
\end{equation}
where $\mathcal{L}$ is the Lagrangian, $\tilde{g}$ is the
determinant of the induced metric on the horizon and
$\hat{\varepsilon}_{ab}$ is the binormal to the horizon.

Using the same prescription as \cite{Myers2010}, we obtain
\begin{align}
\mathcal{S}=S/V_{n-1}&=-2\pi
r_{+}^{n-1}Y \nonumber \\
&=\frac{{r_{0}^{n-1}}}{4}\left(
1+2\,k\hat{\mu}_{2}{\frac{\left(
n-1\right) {L}^{2}}{\left( n-3\right) r_{0}^{2}}}-3k^{2}\hat{\mu}_{3}{\frac{%
\left( n-1\right) {\ L}^{4}}{\left( n-5\right) r_{0}^{4}}}+4{k}\hat{\mu}%
_{4}{\frac{\left( n-1\right) {L}^{6}}{\left( n-7\right)
r_{0}^{6}}}\right)
\end{align}

The temperature of the black holes, after Wick-rotation, is
\begin{equation}
T = \left( \frac{r^{z+1} \sqrt{f^{\prime}g^{\prime}}}{4 \pi
L^{z+1}}\right)_{r=r_0} .
\end{equation}

\begin{figure}[htbp]
    \includegraphics[scale=0.5]{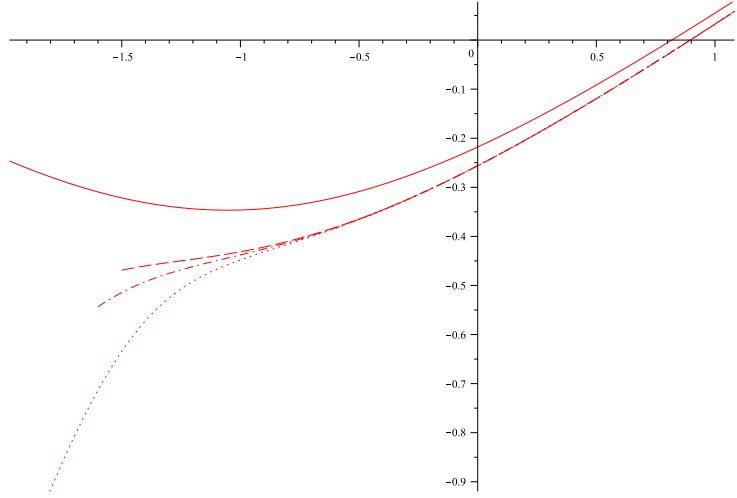}
    \caption{The plot of log($T$) versus log($S$) for higher-curvature black holes,
    where the solid line is Einsteinian, the dashed is Gauss-Bonnet, the dash-dot is $3^{rd}$ order
    quasitopological, and the dotted is $4^{th}$ order (quartic) quasitopological gravity.
    Here, $h_0 = 2.000$, $k=1$, $z=1$, and we are in 4+1 dimensions.
    The other parameters were
    $\hat{\mu}_2=0.04,\hat{\mu}_3=-0.001,\hat{\mu}_4=0.0004$.}
    \label{fig:cvz1}
\end{figure}

We can produce plots to see the effect of the quartic
quasitopological parameter on the thermodynamics of the black
hole. In figure \ref{fig:cvz1}, we see that the positive quartic
parameter continues to act in the same manner as before (i.e. as a
``stronger'' quasitopological addition) by pulling the black hole
solution further from instability.
Recall further that the slope of this graph indicates the sign of
the specific heat, and that in the small-$r$ black hole (leftmost region)
of the Einsteinian case, the
negative slope means that the black hole is unstable.

Note that because the cosmological constant is specified by the
quasitopological parameters, the different solutions do have
different cosmological constants and therefore the plot does not
converge to exactly the same black holes at large $r_+$ (on the
right of the plot). However, if we were to find solutions with the
same cosmological constant we would see that the large black holes
would become thermodynamically identical.
This is the expected result as the larger black holes will have reduced
surface gravity and curvature so the higher order curvature terms
will have reduced effect.

\begin{figure}[htbp]
    \includegraphics[scale=0.5]{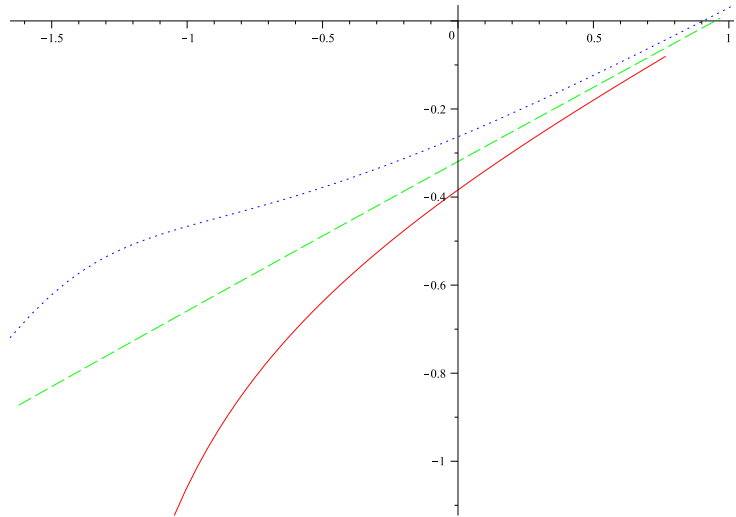}
    \caption{The plot of log($T$) versus log($S$) for higher-curvature black holes,
    where the solid line is $k=-1$, the dashed is $k=0$, the dotted is $k=1$ in
    $4^{th}$ order quasitopological gravity.
    Here, $h_0$ varies slightly for each solution, but is generally around $h_0
    \sim 0.560$, $z=1$, and we are in 4+1 dimensions.
    The other parameters were
    $\hat{\mu}_2=0.04,\hat{\mu}_3=-0.001,\hat{\mu}_4=0.0003$.}
    \label{fig:kvarz1}
\end{figure}

We also plot a comparison of black branes with different hypersurface curvature
in figure \ref{fig:kvarz1}. This shows the expected result that $k=-1$ is
a stable solution while $k=1$ is stable for this particular set of parameters
but appears to have the potential for instability.

\begin{figure}[htbp]
    \includegraphics[scale=0.5]{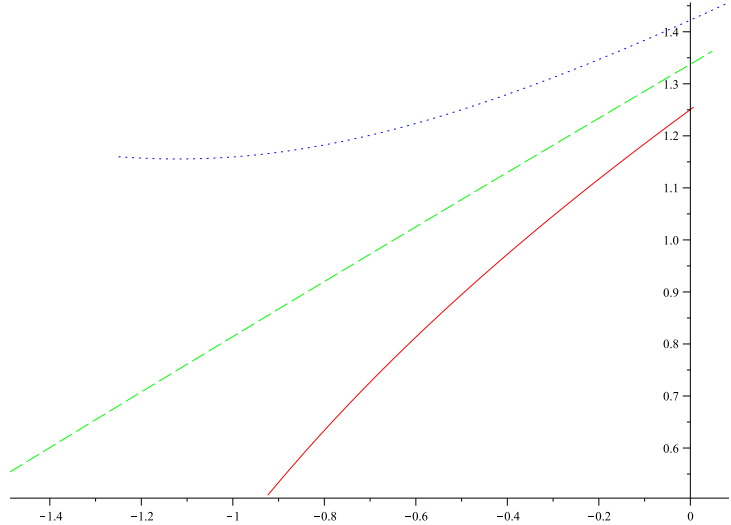}
    \caption{The plot of log($T$) versus log($S$) for higher-curvature black holes,
    where the solid line is $k=-1$, the dashed is $k=0$, the dotted is $k=1$ in
    $4^{th}$ order quasitopological gravity.
    Here, $h_0 = 0.724$, $z=2$, and we are in 4+1 dimensions.
    The other parameters were
    $\hat{\mu}_2=0.04,\hat{\mu}_3=-0.001,\hat{\mu}_4=0.0003$.}
    \label{fig:kvarz2}
\end{figure}

Finally, a thermodynamics plot for $z=2$ was performed, which elucidates
the $z=2$ behaviour of the quartic quasitopological term. Recall that in
$z=2$ cubic quasitopological gravity \cite{Brenna2011}, instabilities in $k=1$
reappear depending on the strength of the cubic coupling term. We see
the same behaviour with the quartic coupling term in figure \ref{fig:kvarz2}.
Though the slope of the curve is difficult to visually ascertain, an examination
of the individual data points yields a transition from a negative to positive slope, as occurs
for some values of $\hat{\mu}_3$ in $z=2$ cubic quasitopological gravity.

\section{Concluding Remarks}
By using the Lifshitz metric in the spherically symmetric case, we
considered the quartic quasitopological gravity and obtained the
field equations.
We then obtained the radially conserved quantity for quartic quasitopological
Lifshitz theories.
We investigated the existence of Lifshitz
solutions both with and without massive background vector field
and we found that a Lifshitz solution can be supported in vacuum
under restrictions on the cosmological constant, Proca mass, and
Proca charge. In the presence of the massive Abelian
gauge field, after demonstrating that the quartic
quasitopological gravity can support a Lifshitz solution, we
numerically derived asymptotically Lifshitz black hole solutions,
comparing them to previously published analytic solutions for consistency.
We found that the $4^{th}-$order quasitopological term acts in a similar
way as the $3^{rd}-$order term on metric functions of the black hole.

We can further summarize our findings for the thermodynamic effect
of the quartic quasitopological parameter. In the context of
cubic quasitopological results, the quartic term does not behave unexpectedly.
Its ability to push solutions towards stability in $z=1$ and to generally
affect stability in $z=2$ was seen.

Our conclusion is that the $4^{th}-$order theory adds yet another nontrivial parameter
to the space of Lifshitz black hole solutions, which may be useful for obtaining
multiple phase transitions, of use in a gauge/gravity duality. Now that
a method of producing numerical solutions has been developed, this space of
thermodynamic behaviour can be more fully explored. We leave this for future work.

\acknowledgments This work has been supported by Payame Noor
University and Jahrom University, as well as by the National
Sciences and Engineering Research Council of Canada. W. B. was
funded by the Vanier CGS Award.

\end{document}